\documentclass{INTERSPEECH2023}
\usepackage{array}
\usepackage{multirow}
\pdfoutput=1 
\newlist{mylist}{enumerate*}{1}
\setlist[mylist]{label=(\roman*)}
\interspeechcameraready

\title{A Neural TTS System with Parallel Prosody Transfer from Unseen Speakers}
\name{Slava Shechtman, Raul Fernandez}
\address{IBM Research AI}
\email{slava@il.ibm.com, fernanra@us.ibm.com}

\begin{document}

\maketitle

\begin{abstract}
Modern neural TTS systems are capable of generating  natural and expressive speech when provided with sufficient amounts of training data. Such systems can be equipped with prosody-control functionality, 
allowing for more direct shaping of the speech output at inference time. 
In some TTS applications, it may be desirable to have an option 
that guides the TTS system with an ad-hoc speech recording exemplar to impose an implicit 
%fine-grained user-defined 
fine-grained, user-preferred
prosodic realization for certain input prompts. In this work we present a first-of-its-kind neural 
%TTS system that is equipped with such an unseen parallel prosody transfer functionality. 
TTS system equipped with such functionality to transfer the prosody from a parallel text recording
from an unseen speaker.
We demonstrate that the proposed system can precisely transfer the speech prosody 
from novel speakers 
to various trained TTS voices with no quality degradation, while preserving the target TTS speakers' identity, as evaluated by a set of subjective listening experiments.       

% It is In the current work we present a neural TTS system learn and transfer desired speaking styles from one seen speaker to another. The also can
% In fine-grained prosody transfer from one speaker to another one aims to mimic the fine speech prosody, while  

\end{abstract}

\section{Introduction}

Over the past decade, Neural Text-to-Speech (TTS) methods have made significant strides in enhancing the naturalness of synthesized speech, utilizing sequence-to-sequence (S2S) architectures ~\cite{Shen-Pang:18, Wang:2017:tacotron, Shen-Jia:20}. 
Originally, those architectures
implicitly predicted speech prosody, and thus lacked prosody controllability. 
Later models were further extended with such functionality. Hierarchical prosodic representations from various temporal scales of speech waveform, either learnt ~\cite{Lee-Kim:19, Sun-Zhang:20} or measured
% SLAVA2Raul 
% removing completely the reference that cite and repeat our work (same granularity) Raitio-Rasipuram:20
~\cite{Shechtman-Sorin:19, Shechtman-Fernandez:21, Shechtman-Fernandez:21b, raitio2022hierarchical}, were recently proposed to manipulate the synthesized  
prosody at various levels of granularity. These representations can be directly obtained from the speech recording, and are designed 
to be used as inputs, along with phoneme sequences, to generate and control high-quality speech output, and then manipulated in an intuitive way at inference time to obtain
a particular prosodic realization~\cite{Shechtman-Sorin:19, Shechtman-Fernandez:21, Shechtman-Fernandez:21b, raitio2022hierarchical}.

In some TTS applications, such as customer-care chat bots, it is useful to have an option to impose a precise user-defined prosody rendering for certain input texts by directly providing an audio recording exemplar to a TTS system. In this setup one expects the system 
to generate a specific prompt while closely mimicking the user's prosody (assuming the same text is spoken by the user) and ignoring the user's identity. We refer to this functionality as \emph{parallel prosody transfer from unseen speakers}, or in short, \emph{Unseen Parallel Prosody Transfer (UPPT)}. In real applications, it is desirable to have not a 
dedicated system to solve this problem, but rather to have an UPPT-enabled architecture
capable of prosody transfer when provided with a \{text, prompt\} pair, and of defaulting to regular inference otherwise when only the text is available.

% have an UPPT-enabled TTS system that supports both the regular TTS inference and the prosody transfer (UPPT). 

Prior work on prosody transfer within S2S models has evolved from 
architectures that transplant broader prosodic features~\cite{SkerryRyan-Battenberg:18}
toward models operating at more fine-grained 
resolutions~\cite{Lee-Kim:19,Klimkov-Ronanki:19,Karlapati-Moinet:20}. One common feature among
these works is that they propose dedicated models that {\em always}
require an exemplar input to generate output, a limitation that we wish to eschew in this
work.
% This REF can go b/c it's only peripherally related to what we do.
%A more recent variant of the problem has looked into providing output guidance 
%by means of high-level textual semantic descriptions~\cite{Guo-Leng:22}, 
%though such work explicitly avoids having to work in the acoustic space which 
%is the problem we solve.
The most similar example from prior art is the approach described in the CopyCat2 
architecture~\cite{Karlapati-Karanasou:22}. This model is both a conventional TTS
system and a prosody-transfer system, and is capable of mimicking the fine-grained prosody  
implicit in a recording. It differs from our work, notably, in that it is confined to
speakers seen in training, and does not accommodate unseen speakers, as
required by the UPPT application.
To the best of our knowledge, the
proposal of an architecture flexible enough to accommodate default synthesis
and prosody transplantation from any arbitrary, unseen speakers while maintaining
high-quality output and preserving the target speaker's identity is
a novel contribution of this work.

\section{Architecture}

When prosodic features are learnt from the training data, special caution should be taken to disentangle the speaker's identity from this prosodic representation, in
order to make it less data sensitive and more suitable to the prosody-transfer task from an unseen speaker~\cite{Karlapati-Karanasou:22}. We have addressed this constraint by adopting a set of Hierarchical Prosody Controls (HPCs)
% TODO: Cite only the original basis of this work, which is ours
%investigated in~\cite{Shechtman-Fernandez:21, raitio2022hierarchical}
% SLAVA2RAUL OK
originally proposed in~~\cite{Shechtman-Fernandez:21}
as an alternative. HPCs are less data sensitive by construction, are extracted
directly from the recordings (no training is involved), and are globally normalized,
making them only weakly dependent on the training data (due to the normalization procedure~\cite{Shechtman-Sorin:19}) and more suitable to the UPPS use case. It was 
previously shown that HPCs are speaker agnostic~\cite{Shechtman-Fernandez:21b}, are able to generate a wide variety of speaking 
% TODO: Reduce the citing here as well,and in the next paragraph?
% SLAVA2RAUL YES
styles~\cite{Shechtman-Fernandez:21b, Fernandez-Shechtman:22}, and provide word-level % emphasis control~\cite{Shechtman-Fernandez:21, raitio2022hierarchical}.
emphasis control~\cite{Shechtman-Fernandez:21}.

\begin{figure}[t]
  \centering
  \includegraphics[width=\linewidth]{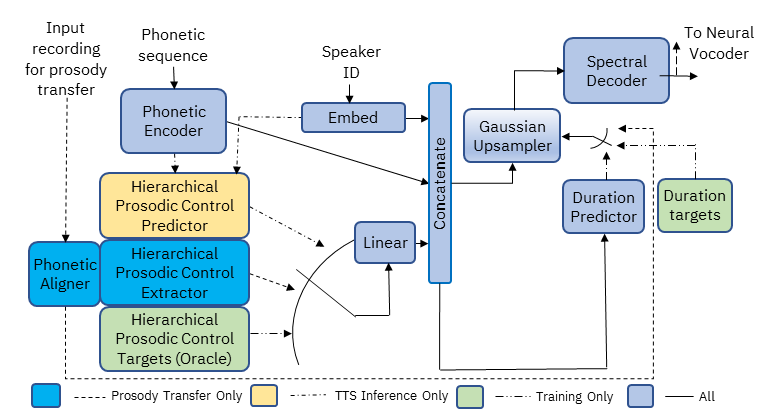}
  \setlength{\abovecaptionskip}{-8pt}
  \setlength{\belowcaptionskip}{-12pt}
  \caption{
  The proposed HPC-controllable neural TTS with a prosody transfer capability}
 \vspace{-2mm}
  \label{fig:arch}
\end{figure}

In this work we precisely follow the
% % SLAVA2RAUL remove here raitio2022hierarchical
HPC-controllable~\cite{Shechtman-Fernandez:21} Non-Attentive-Tacotron2 (NAT2)~\cite{Shen-Jia:20} architecture proposed in~\cite{Fernandez-Shechtman:22} as the speech-generation module of our system (see Fig.~\ref{fig:arch}). We hypothesize, though, that the prosody-transfer precision can benefit from 
HPCs with a finer resolution than those proposed in those works, 
and consequently experiment with various kinds of HPC hierarchies, as described in Section~\ref{sec:HPC}.   
As detailed in~\cite{Fernandez-Shechtman:22}, an input phonetic sequence passes through a Phonetic Encoder, is combined with the embedded HPC feature sequence, undergoes upsampling based on the predicted phone durations, and generates a sequence of acoustic feature vectors by an autoregressive Spectral Decoder that are finally fed to an independently trained LPCNet~\cite{Valin-Skoglund:19} neural vocoder (see Fig.~\ref{fig:arch}).

During training, the acoustic decoder  obtains the ground-truth HPCs and minimizes the acoustic regression loss, based on $L_1 + L_2$ loss operator~\cite{Shen-Pang:18}, plus the duration prediction loss~\cite{Shen-Jia:20}. During inference, the system supports two modes of operation: a TTS mode and a prosody-transfer mode. To enable the TTS mode, an HPC predictor module is separately trained~\cite{Fernandez-Shechtman:22}. The HPC predictor is fed with the pretrained phonetic encoder (with all its trainable weights frozen) and predicts the HPC sequence by minimizing an MSE regression loss~\cite{Shechtman-Fernandez:21b}. 
To enable the prosody-transfer mode, one has to provide a parallel audio recording in addition to the regular TTS input. The HPC parameter sequence is extracted directly from the recording (as detailed in section~\ref{sec:TBE}) and fed to the speech generation model, bypassing the HPC sequence prediction. While in prosody-transfer mode, the phone durations from the input recording can be either directly used or predicted by the duration predictor.

\subsection{Hierarchical Prosody Controls (HPCs)}
\label{sec:HPC}
Hierarchical Prosody Controls (HPCs) comprise a low-dimensional set of temporal prosodic measurements (e.g. rhythm, pitch, energy), evaluated hierarchically over several linguistically-meaningful temporal intervals~\cite{Shechtman-Fernandez:21b}, e.g., sentence- and word-intervals~\cite{Shechtman-Fernandez:21b} or utterance- and phone-intervals~\cite{raitio2022hierarchical}.

In this work we make use of a set of four prosodic measurements~\cite{Shechtman-Fernandez:21b} at various temporal hierarchies, as summarized below. In addition to the sentence- and  word-level HPCs~\cite{Shechtman-Fernandez:21b}, we investigate here how adding finer HPC granularity (i.e. syllable- and phone-level features) can facilitate the prosody-transfer precision. Therefore, we generalize the HPC formulation in ~\cite{Shechtman-Fernandez:21b} to support any amount of temporal hierarchy levels.  

Let $T$ be a desired HPC temporal hierarchy, e.g., $T=(sentence, word, syllable)$. 
Then, the absolute prosodic measurement set $H^{T_i}=\{h^{T_i}_{dur},h^{T_i}_{{\Delta}f_0},h^{T_i}_{f_0},h^{T_i}_{{\angle}f_0}\}$ is evaluated (per intervals $T_i$), as follows: 

\begin{itemize}
\item $h^{T_i}_{dur}$: The log of the average phone duration, along $T_i$.
\item $h^{T_i}_{{\Delta}f_0}$: The $f_0$ dynamics (i.e., the difference between the 95- and 5-percentiles of log-$f_0$), along $T_i$.
\item $h^{T_i}_{f_0}$: The median log-$f_0$, along $T_i$ minus the median log-$f_0$, along a corresponding single speaker data set. Note that the second term is required to make the absolute pitch measurement gender-agnostic.
\item $h^{T_i}_{{\angle}f_0}$: The log-$f_0$ linear regression slope along $T_i$.
\end{itemize}

The $f_0$-based HPC measurements are performed based on a pitch trajectory obtained by the \emph{RAPT} pitch detector~\cite{talkin1995robust} at 5ms time steps, and linearly interpolated 
through unvoiced regions.

The previous prosodic measurements also require phonetic alignment of the input 
waveform (at least for the fine-grained prosodic measurements). While offline forced alignment is not usually a problem for high-quality TTS corpora, the precision of phonetic alignment can deteriorate in the UPPT use case, where just a single utterance of unknown audio quality from an unseen speaker 
must be phonetically aligned (see Section~\ref{sec:TBE}).    

The $H^{T_i}$ measurements are performed once per $T_i$ interval and then propagated down to the
temporal granularity of the phonetic encoder outputs (i.e., phones)
to form piecewise functions that are constant within the corresponding $T_i$ intervals.
Based on the propagated absolute interval measurements $\widetilde{H}^{T_i}$ we construct the unnormalized  \emph{hierarchical} measurements by concatenating the corresponding \emph{residual} interval measurements, according to (\ref{eq:prsinfo}).
\begin{equation}
P = [\widetilde{H}^{T_0}, \widetilde{H}^{T_1}-\widetilde{H}^{T_0},..., \widetilde{H}^{T_i}-\widetilde{H}^{T_{i-1}},...]
\label{eq:prsinfo}
\end{equation}

Eventually, each input utterance is represented by a {\em normalized} HPC matrix $\hat{P}$ of size $N_{phones}\times N_{HPC}$ (with the $k$-th column of $\hat{P}$ denoted by $\hat{p}_k$),  
obtained by applying a global (corpus-wise) normalization on $P$ (with $p_k$ as its $k$-th column). 
Let the $k$-th component's global mean and standard deviation be $\mu_k$ and $\sigma_k$ 
respectively. The normalization is then given by
\begin{equation}
\hat{p}_k = (p_k - \mu_k)/3\sigma_k
\label{eq:prsinfo_norm}
\end{equation}  
In this work we explore the TTS and UPPT modes based on $(sentence, word)$, $(sentence, word, syllable)$ and $(sentence, word, phone)$ hierarchies.   
% \begin{itemize}
% \item $T^{ref}=(sentence, word)$ with $4\times2=8$ HPC features per phone, as in~\cite{Shechtman-Fernandez:21b},   
% \item $T^{syl}=(sentence, word, syllable)$ with $4\times3=12$ HPC features per phone,
% \item $T^{phn}=(sentence, word, phone)$ with $4\times3=12$ HPC features per phone.
% \end{itemize}

To support inference in TTS mode when observed HPC targets are unavailable, an HPC predictor is
trained to generate the appropriate measurements. We empirically found that the original HPC predictor architecture (i.e., 3 stacked Bi-LSTM layers with 128 hidden nodes with output linear layer)~\cite{Shechtman-Fernandez:21b} works well for the proposed HPC variants. Large-scale crowd-sourced MOS tests 
(up to 100 subjects, 40 samples per system, 25 votes per sample) revealed that an HPC-controlled \textit{NAT2} architecture~\cite{Fernandez-Shechtman:22} generates speech with the same quality and naturalness for all the proposed variants.  

\subsection{Unseen Parallel Prosody Transfer (UPPT)}
\label{sec:TBE}

We consider two approaches for HPC-based prosody transfer within a NAT2 TTS architecture~\cite{Fernandez-Shechtman:22} customized for UPPT:
\begin{itemize}
\item \emph{HPC import}: the sequence of HPC features is extracted from the input audio, the duration predictor module is further applied to predict the phone durations. In that case the rhythm transfer may become less precise, but this setup is less vulnerable to alignment errors.    
\item \emph{HPC and duration import}: the sequence of HPC features and phone durations are directly extracted from the input audio. Timing transfer may be more precise but also more sensitive to potential alignment errors.     
\end{itemize}

Both approaches require automatic alignment of a single utterance of potentially noisy quality
and uttered by unseen speakers. This can be a challenge, and certain alignment errors are inevitable, so there is a certain trade-off between the {\em average} prosody transfer precision (depending on the HPC granularity) and the amount of perceivable {\em local} quality issues due to occasional alignment errors.
%Luckily, an off-the-shelf open source solution was found suitable for our purposes. 
In our work we make use of the open-source Montreal Forced Aligner (MFA) package~\cite{mcauliffe17_interspeech} 
with its pretrained US English triphone acoustic model \emph{english\_us\_arpa} and default speaker adaptation~\cite{mcauliffe17_interspeech} to obtain the temporal intervals for HPC calculations. Although rare alignment problems resulted in occasional local audible quality deterioration, we found it performs reasonably well on unseen data of various quality (see more details in Section~\ref{sec:eval}). 

When evaluating HPCs (see Section~\ref{sec:HPC}) for an utterance from a novel speaker, one needs to estimate their median log-$f_0$ to evaluate the $h^{T_i}_{f_0}$ components. We do this based on the input utterance. Additionally, for the final normalization we use the pre-stored global (multi-speaker) statistics collected from the seen speakers (i.e., the TTS  training data).

%TODO:  elaborate on the HPC loss if we have space

% This architecture comb non-attentive Tacotron2~\cite{Shen-Pang:18} architecture (NAT2) from~\cite{Shen-Jia:20}, augmented with the hierarchical prosodic controls  (HPCs)~\cite{Shechtman-Fernandez:21,Shechtman-Fernandez:21b}. 

% We hypothesize that the hierarchical prosody control features~\cite{Shechtman-Fernandez:21b, raitio2022hierarchical} can be suitable also for the prosody transfer task, however in this work we'd like to explore fine-grained prosody controls more fine-grained temporal hierarchies are required to gain more precise prosody transfer HPC-controlled non-attentive Tacotron Our model 

\section{Experimental Setup}

The training material for our speech synthesis system comprises proprietary wide-band (22.05 kHz) speech corpora, ranging from 16k to 23k sentences, from three professional native speakers of US English (2 females and 1 male) uttered in various speaking styles.
One female and one male voice were selected  as target synthesis voices to evaluate prosody transplantation in a \emph{neutral} style.
For UPPT evaluation we constructed a set of 40 source utterances containing versatile samples from various speakers, speaking styles and data sets, selected by listening to the reference recordings only, according
to the following guidelines: Choose samples between 5 and 25 words that convey ``interesting'' (``non-average'') prosodic patterns, and try to avoid many long sentences, as that might make the subjective
assessment task more difficult for the listeners.
%\begin{mylist}
%  \item the samples contain between 5 and 25 words,
%  \item the samples convey "interesting" (i.e., "non-average") prosodic patterns, and
%  \item the shorter samples are preferred to ease the subjective assessment by crowd.
%\end{mylist}
Based on this, we selected the following data for six speakers:
\begin{itemize}
\item 20 \emph{out-of-domain}, unseen samples from the \emph{test-clean} section of LibriTTS~\cite{zen2019libritts}, uttered by four \emph{unseen} unprofessional speakers (two females (set L-F) and two males (set L-M); five utterances from each). The selected speaker IDs are: \emph{121\_127105}, \emph{1089\_134686}, \emph{2300\_131720}, \emph{3575\_170457}. 
\item 10 \emph{in-domain} unseen recordings from one \emph{unseen} professional female speaker
\item 10 \emph{in-domain} unseen recordings from one \emph{seen} (in training) professional male speaker
\end{itemize}

We trained the following multi-speaker TTS systems sharing the prosody-controlled NAT2 architecture of~\cite{Fernandez-Shechtman:22} with various prosody controls to assess different prosody transplantation techniques. (See Section~\ref{sec:TBE} for more details on the HPC definitions and durations employed by Systems 2-8.)

\begin{enumerate}
\item \emph{Ref}: a system that implements classic prosody transfer by means of \emph{reference encoding}~\cite{SkerryRyan-Battenberg:18} instead of HPCs. Here the trainable \emph{reference encoder}
generates a fixed-sized utterance-level prosodic embedding from the input spectrogram. This embedding is broadcast-concatenated with the phonetic encoder outputs~\cite{SkerryRyan-Battenberg:18} (instead of HPCs in Fig.~\ref{fig:arch}) and then fed into the spectral decoder. For the sake of consistency, the remaining encoder-decoder architecture is identical to the rest of the HPC-controlled systems.

\item \emph{HPC0-TTS}: an HPC-controlled TTS system~\cite{Fernandez-Shechtman:22}, that deploys the two-level HPCs~\cite{Fernandez-Shechtman:22, Shechtman-Fernandez:21b} (sentence- and word- level). On inference, the HPCs are predicted from the input phonetic sequence. This system does no prosody transfer.

\item \emph{HPC0-D0}: an HPC-controlled prosody transfer system~\cite{Fernandez-Shechtman:22}, that deploys the two-level HPCs~\cite{Fernandez-Shechtman:22, Shechtman-Fernandez:21b} (sentence- and word- level) and applies \emph{HPC import} only.

%\item \emph{HPC0-D1}: an HPC-controlled prosody transfer system~\cite{Fernandez-Shechtman:22}, that %deploys the two-level HPCs~\cite{Fernandez-Shechtman:22, Shechtman-Fernandez:21b} (sentence- and word- %level) and applies \emph{HPC and duration import} (see Section \ref{sec:TBE})  
\item \emph{HPC0-D1}: Like \emph{HPC0-D0} with additional \emph{duration import}.

\item \emph{HPC1-D0}: an HPC-controlled prosody transfer system~\cite{Fernandez-Shechtman:22}, that deploys three-level HPCs (sentence-, word- and syllable- level) and applies \emph{HPC import} only.
%\item \emph{HPC1-D1}: an HPC-controlled prosody transfer system~\cite{Fernandez-Shechtman:22}, that %deploys three-level HPCs (sentence-, word- and syllable- level) and applies \emph{HPC and duration %import} (see Section \ref{sec:TBE})  
\item \emph{HPC1-D1}: Like \emph{HPC1-D0} with additional \emph{duration import}.

\item \emph{HPC2-D0}: an HPC-controlled prosody transfer system~\cite{Fernandez-Shechtman:22}, that deploys three-level HPCs (sentence-, word- and phone- level) and applies \emph{HPC import} only.
%\item \emph{HPC2-D1}: an HPC-controlled prosody transfer system~\cite{Fernandez-Shechtman:22}, that %deploys three-level HPCs (sentence-, word- and phone- level) and applies \emph{HPC and duration %import} (see Section \ref{sec:TBE})  
\item \emph{HPC2-D1}:  Like \emph{HPC2-D0} with additional \emph{duration import}.
\end{enumerate}

\begin{table}[h]
  \setlength{\belowcaptionskip}{-8pt}
  \setlength{\abovecaptionskip}{2pt}
  \caption{4-scale prosody similarity after the prosody transfer to a male target speaker} 
  \label{tab:male_prossim}
  \footnotesize
  \centering
  \begin{tabular}{cc|cc|cc|c|m{1em}} \toprule
    & & Dissim.$^+$ & Dissim. & Sim. & Sim.$^+$ & Score & Rank\\ \midrule
    
    \multirow{3}{1em}{Ref.} & All & 40.8\%  & 31.5\% & 20.0\% & 7.8\% & 1.95 & 5\\
    % 28.50%	39.00%	23.00%	9.50%	200.00	135	65	2.135
    & L-F & 28.5\%  & 39.0\% & 23.0\% & 9.5\% & 2.14 & \\ 
    % 47.00%	28.50%	17.50%	7.00%	200.00	151	49	1.845
    & L-M & 47.0\%  & 28.5\% & 17.5\% & 7.0\% & 1.85 & \\ \midrule 
    
    % HPC0-TTS & 38.1\%  & 27.6\% & 23.6\% & 10.6\% & 2.065 & 4\\ \midrule
    \multirow{3}{1em}{HPC0-TTS} & All & 38.1\%  & 27.6\% & 23.6\% & 10.6\% & 2.07 & 4\\
    % 30.50%	29.50%	30.00%	10.00%	200.00	120	80	2.195
    & L-F & 30.5\%  & 29.5\% & 30.0\% & 10.0\% & 2.20 & \\ 
    % 43.00%	24.50%	22.50%	10.00%	200.00	135	65	1.995
    & L-M & 43.0\%  & 24.5\% & 22.5\% & 10.0\% & 2.00 & \\ \midrule
    
    % HPC0-D0 & 9.1\% & 19.5\% & 38.1\% & 33.2\% & 2.952 & 3\\ \midrule
    \multirow{3}{1em}{HPC0-D0} & All & 9.1\% & 19.5\% & 38.1\% & 33.2\% & 2.95 & 3\\ 
    % 13.50%	21.50%	37.00%	28.00%	200.00	70	130	2.795
    & L-F & 13.5\%  & 21.5\% & 37.0\% & 28.0\% & 2.80 & \\ 
    % 8.50%	22.50%	40.00%	29.00%	200.00	62	138	2.895
    & L-M & 8.5\%  & 22.5\% & 40.0\% & 29.0\% & 2.90 & \\ \midrule
    
    % HPC0-D1 & 7.2\% & 17.2\% & 34.4\% & 41.1\% & 3.092 & 3\\ \midrule
    \multirow{3}{1em}{HPC0-D1} & All & 7.2\% & 17.2\% & 34.4\% & 41.1\% & 3.09 & 3\\
    % 9.00%	28.00%	32.00%	31.00%	200.00	74	126	2.85
    & L-F & 9.0\%  & 28.0\% & 32.0\% & 31.0\% & 2.85 & \\ 
    % 10.00%	16.00%	34.50%	39.50%	200.00	52	148	3.035
    & L-M & 10.0\%  & 16.0\% & 34.5\% & 39.5\% & 3.04 & \\ \midrule
    
    % HPC1-D0 & 8.5\% & 17.6\% & 37.5\% & 36.4\% & 3.018 & 3\\ \midrule
    \multirow{3}{1em}{HPC1-D0} & All & 8.5\% & 17.6\% & 37.5\% & 36.4\% & 3.02 & 3\\
    % 13.50%	20.00%	36.50%	30.00%	200.00	67	133	2.83
    & L-F & 13.5\%  & 20.0\% & 36.5\% & 30.0\% & 2.83 & \\ 
    % 6.50%	19.50%	35.00%	39.00%	200.00	52	148	3.065
    & L-M & 6.5\%  & 19.5\% & 35.0\% & 39.0\% & 3.07 & \\ \midrule
    
    % HPC1-D1 & 5.2\% & 14.8\% & 38.4\% & 41.6\% & 3.164 & 2\\ \midrule
    \multirow{3}{1em}{HPC1-D1} & All & 5.2\% & 14.8\% & 38.4\% & 41.6\% & 3.16 & 2\\
    % 8.50%	17.00%	35.00%	39.50%	200.00	51	149	3.055
    & L-F & 8.5\%  & 17.0\% & 35.0\% & 39.5\% & 3.06 & \\ 
    % 5.50%	11.50%	40.50%	42.50%	200.00	34	166	3.2
    & L-M & 5.5\%  & 11.5\% & 40.5\% & 42.5\% & 3.2 & \\ \midrule
    
    % HPC2-D0 & 4.1\% & 11.9\% & 38.2\% & 45.8\% & 3.257 & 1\\ \midrule
    \multirow{3}{1em}{HPC2-D0} & All & 4.1\% & 11.9\% & 38.2\% & 45.8\% & 3.257 & 1\\
    % 6.50%	14.50%	42.00%	37.00%	200.00	42	158	3.095
    & L-F & 6.5\%  & 14.5\% & 42.0\% & 37.0\% & 3.10 & \\ 
    % 2.50%	11.00%	33.50%	53.00%	200.00	27	173	3.37
    & L-M & 2.5\%  & 11.0\% & 33.5\% & 53.0\% & 3.37 & \\ \midrule
    
    % HPC2-D1 & 5.4\% & 11.5\% & 36\% & 47.1\% & 3.248 & 1\\ 
    \multirow{3}{1em}{HPC2-D1} & All & 5.4\% & 11.5\% & 36\% & 47.1\% & 3.248 & 1\\
    % 8.50%	17.00%	34.50%	40.00%	200.00	51	149	3.06
    & L-F & 8.5\%  & 17.0\% & 34.5\% & 40.0\% & 3.06 & \\ 
    % 4.00%	10.00%	36.00%	50.00%	200.00	28	172	3.32
    & L-M & 4.0\%  & 10.0\% & 36.0\% & 50.0\% & 3.32 & \\ 

    \bottomrule
  \end{tabular}
\vspace{-4mm}
\end{table}

\begin{table}[h]
  \setlength{\belowcaptionskip}{-8pt}
  \setlength{\abovecaptionskip}{2pt}
  \caption{4-scale prosody similarity after the prosody transfer to a female target speaker} 
  \label{tab:female_prossim}
  \footnotesize
  \centering
  \begin{tabular}{cc|cc|cc|c|m{1em}} \toprule
    & & Dissim.$^+$ & Dissim. & Sim. & Sim.$^+$ & Score & Rank\\ \midrule
    
    %     Ref & 38.1\%  & 31.9\% & 21.1\% & 8.9\% & 2.008 & 4\\ \midrule    
    \multirow{3}{1em}{Ref.} & All & 38.1\%  & 31.9\% & 21.1\% & 8.9\% & 2.01 & 4\\
    % 33.50%	31.50%	24.50%	10.50%	200.00	130	70	2.12
    & L-F & 33.5\%  & 31.5\% & 24.5\% & 10.5\% & 2.12 & \\ 
    % 34.50%	32.00%	21.00%	12.50%	200.00	133	67	2.115
    & L-M & 34.5\%  & 32.0\% & 21.0\% & 12.5\% & 2.12 & \\ \midrule 
    
    %     HPC0-TTS & 38.1\%  & 29.6\% & 23.8\% & 8.5\% & 2.027 & 4\\ \midrule
    \multirow{3}{1em}{HPC0-TTS} & All & 38.1\%  & 29.6\% & 23.8\% & 8.5\% & 2.03 & 4\\
    % 33.50%	27.00%	30.50%	9.00%	200.00	121	79	2.15
    & L-F & 33.5\%  & 27.0\% & 30.5\% & 9.0\% & 2.15 & \\ 
    % 35.50%	29.50%	24.00%	11.00%	200.00	130	70	2.105
    & L-M & 35.5\%  & 29.5\% & 24.0\% & 11.0\% & 2.11 & \\ \midrule
    
    %     HPC0-D0 & 8.5\% & 21.8\% & 38.1\% & 31.6\% & 2.928 & 3\\ \midrule
    \multirow{3}{1em}{HPC0-D0} & All & 8.5\% & 21.8\% & 38.1\% & 31.6\% & 2.93 & 3\\ 
    % 11.50%	24.50%	34.50%	29.50%	200.00	72	128	2.82
    & L-F & 11.5\%  & 24.5\% & 34.5\% & 29.5\% & 2.82 & \\ 
    % 7.00%	21.00%	38.00%	34.00%	200.00	56	144	2.99
    & L-M & 7.0\%  & 21.0\% & 38.0\% & 34.0\% & 2.99 & \\ \midrule
    
    %     HPC0-D1 & 6.2\% & 16.6\% & 40.1\% & 37.0\% & 3.077 & 2\\ \midrule
    \multirow{3}{1em}{HPC0-D1} & All & 6.2\% & 16.6\% & 40.1\% & 37.0\% & 3.08 & 2\\
    % 6.50%	20.00%	32.50%	41.00%	200.00	53	147	3.08
    & L-F & 6.5\%  & 20.0\% & 32.5\% & 41.0\% & 3.08 & \\ 
    % 9.00%	15.50%	36.50%	39.00%	200.00	49	151	3.055
    & L-M & 9.0\%  & 15.5\% & 36.5\% & 39.0\% & 3.08 & \\ \midrule
    
    %     HPC1-D0 & 5.8\% & 15.9\% & 40.1\% & 38.2\% & 3.107 & 2\\ \midrule
    \multirow{3}{1em}{HPC1-D0} & All & 5.8\% & 15.9\% & 40.1\% & 38.2\% & 3.11 & 2\\
    % 8.50%	20.00%	37.00%	34.50%	200.00	57	143	2.975
    & L-F & 8.5\%  & 20.0\% & 37.0\% & 34.5\% & 2.98 & \\ 
    % 5.00%	17.50%	34.00%	43.50%	200.00	45	155	3.16
    & L-M & 5.0\%  & 17.5\% & 34.0\% & 43.5\% & 3.16 & \\ \midrule
    
    %     HPC1-D1 & 5.1\% & 11.2\% & 37.6\% & 46\% & 3.243 & 1\\ \midrule
    \multirow{3}{1em}{HPC1-D1} & All & 5.1\% & 11.2\% & 37.6\% & 46\% & 3.24 & 1\\
    % 6.50%	11.50%	39.00%	43.00%	200.00	36	164	3.185
    & L-F & 6.5\%  & 11.5\% & 39.0\% & 43.0\% & 3.19 & \\ 
    % 5.50%	10.50%	40.50%	43.50%	200.00	32	168	3.22
    & L-M & 5.5\%  & 10.5\% & 40.5\% & 43.5\% & 3.22 & \\ \midrule
    
    %     HPC2-D0 & 3.8\% & 11.8\% & 36.2\% & 48.2\% & 3.288 & 1\\ \midrule
    \multirow{3}{1em}{HPC2-D0} & All & 3.8\% & 11.8\% & 36.2\% & 48.2\% & 3.288 & 1\\
    % 3.50%	14.00%	37.50%	45.00%	200.00	35	165	3.24
    & L-F & 3.5\%  & 14.0\% & 37.5\% & 45.0\% & 3.24 & \\ 
    % 2.50%	13.00%	34.50%	50.00%	200.00	31	169	3.32
    & L-M & 2.5\%  & 13.0\% & 34.5\% & 50.0\% & 3.32 & \\ \midrule
    
    %     HPC2-D1 & 3.8\% & 12.2\% & 35.6\% & 48.4\% & 3.286 & 1\\ 
    \multirow{3}{1em}{HPC2-D1} & All & 3.8\% & 12.2\% & 35.6\% & 48.4\% & 3.286 & 1\\
    % 5.00%	15.50%	35.50%	44.00%	200.00	41	159	3.185
    & L-F & 5.0\%  & 15.5\% & 35.5\% & 44.0\% & 3.19 & \\ 
    % 2.50%	10.50%	33.50%	53.50%	200.00	26	174	3.38
    & L-M & 2.5\%  & 10.5\% & 33.5\% & 53.5\% & 3.38 & \\ 
    \bottomrule
  \end{tabular}
\vspace{-9mm}
\end{table}

\section{Evaluation}
\label{sec:eval}

We designed a set of subjective evaluations to assess how well the prosody is transferred from various input utterances to a male and a female target voices that are part of a multi-speaker TTS training corpus. We conducted several crowd-based subjective listening tests to evaluate :
\begin{mylist}
  \item prosody similarity,
  \item quality \& naturalness,
  \item speaker similarity.
\end{mylist}
All experiments were conducted on the AMT crowd-sourcing platform with votes collected from 30-45 subjects qualified as \emph{masters}~\cite{sodre2017analysis}. 40 parallel stimuli with identical texts per system were used in the \emph{prosody similarity} and \emph{quality \& naturalness} tests with each stimuli assessed by 20 distinct subjects on average (800 votes per system). \emph{Speaker similarity} was tested with 10 stimuli per system, assessed by 20 distinct speakers each (200 votes per system). The outcomes for each target speaker (male and female) were evaluated in distinct experiments.

% The prosody transfer -- This is referred to by various names. Choosing one for consistency -- RF 
{\em (i) Prosody similarity} was assessed by a 4-level pairwise similarity test, as in~\cite{wester16_interspeech}, where subjects assessed unordered stimuli pairs with one stimulus containing an input recording with the source prosody, and the other 
a corresponding synthetic sample uttered by either of the prosody transfer systems (randomized). The subjects were asked to "ignore the speaker identity" and "judge how similar they find the samples in terms of \emph{how the speakers are saying them}, i.e., their intonation, speaking pace, rhythm, pausing, etc." The 4-level scale was labeled with "Very dissimilar" (Dissim$^+$), "Somewhat Dissimilar" (Dissim), "Somewhat similar" (Sim), "Very similar" (Sim$^+$). The results for this test are presented in Tables~\ref{tab:male_prossim} and~\ref{tab:female_prossim},
showing the distribution over raw similarity values plus the average score (assuming values 1 to 4 in the 4-level categorical scale). A Barnard’s exact test~\cite{barnard1945new} (two-sided, $p=0.05$) was used to calculate significance between systems on the binary similar/dissimilar votes to determine 
a ranking among the 8 systems (or among groupings thereof) that do not differ significantly from each other in terms of prosodic similarity. We are including in the rightmost column the rank received by the system from column 1 (smallest is best). In addition to the results for \emph{all} the UPPT stimuli set (6 speakers), we also present the scores for the more challenging input subsets (i.e. unprofessional LibriTTS~\cite{zen2019libritts}), pooled by gender. These results 
%in Tables \ref{tab:male_prossim} and \ref{tab:female_prossim} 
demonstrate that the proposed HPC-based systems (of various HPC granularity) significantly outperform the reference system. The results also reveal that adding finer granularity to HPC features as well as importing phone durations gradually improve the prosody transfer precision. There is a high variance for the perceived prosody similarity when transferring prosody from various input voices. However, one can clearly notice that the same unseen voices perform similarly for both same-gender and cross-gender prosody transfer. One can also observe that prosody transfer from out-of-domain unprofessional recordings is not consistently worse than the overall performance (that includes professional recordings of the in-domain material).

% \begin{table}[h]
%   \setlength{\belowcaptionskip}{-8pt}
%   \setlength{\abovecaptionskip}{2pt}
%   \caption{Prosody similarity after the prosody transfer to a female target speaker} 
%   \label{tab:female_prossim}
%   \footnotesize
%   \centering
%   \begin{tabular}{c|cc|cc|c|c} \toprule
%     & Dissim.$^+$ & Dissim. & Sim. & Sim.$^+$ & Score & Rank \\ \midrule
%     % sp0_REFENC	38.10%	31.90%	21.10%	8.90%	800	560	240	2.008
%     Ref & 38.1\%  & 31.9\% & 21.1\% & 8.9\% & 2.008 & 4\\ \midrule
%     % sp0_NEU	38.10%	29.60%	23.80%	8.50%	800	542	258	2.027
%     HPC0-TTS & 38.1\%  & 29.6\% & 23.8\% & 8.5\% & 2.027 & 4\\ \midrule
%     % sp0_101_D0	8.50%	21.80%	38.10%	31.60%	800	242	558	2.928
%     HPC0-D0 & 8.5\% & 21.8\% & 38.1\% & 31.6\% & 2.928 & 3\\ \midrule
%     % sp0_101_D1	6.20%	16.60%	40.10%	37.00%	800	182	618	3.077
%     HPC0-D1 & 6.2\% & 16.6\% & 40.1\% & 37.0\% & 3.077 & 2\\ \midrule
%     % sp0_10110_D0	5.80%	15.90%	40.10%	38.20%	800	174	626	3.107
%     HPC1-D0 & 5.8\% & 15.9\% & 40.1\% & 38.2\% & 3.107 & 2\\ \midrule
%     % sp0_10110_D1	5.10%	11.20%	37.60%	46.00%	800	130	670	3.243
%     HPC1-D1 & 5.1\% & 11.2\% & 37.6\% & 46\% & 3.243 & 1\\ \midrule
%     % sp0_10111_D0	3.80%	11.80%	36.20%	48.20%	800	125	675	3.288
%     HPC2-D0 & 3.8\% & 11.8\% & 36.2\% & 48.2\% & 3.288 & 1\\ \midrule
%     % sp0_10101_D1	3.80%	12.20%	35.60%	48.40%	800	128	672	3.286
%     HPC2-D1 & 3.8\% & 12.2\% & 35.6\% & 48.4\% & 3.286 & 1\\ 

%     \bottomrule
%   \end{tabular}
% \vspace{-2mm}
% \end{table}

\begin{table}[t]
  \setlength{\belowcaptionskip}{-8pt}
  \setlength{\abovecaptionskip}{2pt}
  \caption{MOS scores and alignment error count after the prosody transfer 
  % (separate experiments for a male and a female voices) 
  }
  \label{tab:mos}
  \footnotesize
  \centering
  \begin{tabular}{c|cc|cc} \toprule
   \multirow{2}{*}{System} & \multicolumn{2}{c|}{F} & \multicolumn{2}{c}{M} \\ \cmidrule{2-5} 
           & MOS & Aln.Err. & MOS & Aln.Err. \\ \midrule
%   \multirow{2}{*}{F1} & 3.93 & 3.84 & 3.82 & 4.00 \\ 
%                       & (.86) & (.85) & (.86) &  (.86) \\ \midrule
  Ref. & 3.62 ± 0.07 & 0 & 3.46 ± 0.07 & 0  \\ \midrule
  HPC0-TTS & 3.86 ± 0.06 & 0 & 3.71 ± 0.07 & 0  \\ \midrule
  HPC0-D0 & 3.89 ± 0.06 & 0 & 3.75 ± 0.07 & 0  \\ \midrule
  HPC0-D1 & 3.87 ± 0.07 & 0 & 3.71 ± 0.07 & 1  \\ \midrule
  HPC1-D0 & 3.94 ± 0.06 & 0 & 3.74 ± 0.07 & 1  \\ \midrule
  HPC1-D1 & 3.94 ± 0.06 & 3 & 3.77 ± 0.07 & 1  \\ \midrule
  HPC2-D0 & 3.88 ± 0.07 & 5 & 3.74 ± 0.07 & 3  \\ \midrule
  HPC2-D2 & 3.92 ± 0.06 & 7 & 3.74 ± 0.07 & 2  \\ \midrule
  PCM & 4.51 ± 0.05 & 0 & 4.51 ± 0.05 & 0  \\ 
    \bottomrule
  \end{tabular}
\end{table}

\begin{table}[h]
  \setlength{\belowcaptionskip}{-8pt}
  \setlength{\abovecaptionskip}{2pt}
  \caption{4-scale speaker similarity after the prosody transfer compared to the TTS inference  % (distinct tests for a female and a male target speaker)
  } 
  \label{tab:spksim}
  \footnotesize
  \centering
  \begin{tabular}{cc|cc|cc|c|m{1em}} \toprule
    & & Dissim.$^+$ & Dissim. & Sim. & Sim.$^+$ & Score & Rank\\ \midrule
    % 17.10%	22.10%	27.10%	33.70%	199.00	78	121	2.774
    \multirow{2}{1em}{Ref} & F & 17.1\%  & 22.1\% & 27.1\% & 33.7\% & 2.77 & 3\\
    % 12.00%	17.80%	41.40%	28.80%	192.00	57	135	2.87	341.4
    & M & 12.0\%  & 17.8\% & 41.4\% & 28.8\% & 2.87 & 3\\ \midrule
    
    % 5.00%	12.10%	31.70%	51.30%	199.00	34	165	3.295	2
    \multirow{2}{2em}{HPC0-D0} & F & 5.0\%  & 12.1\% & 31.7\% & 51.3\% & 3.30 & 2\\
    % 2.60%	12.60%	38.70%	46.10%	192.00	29	163	3.283	2
    & M & 2.6\%  & 12.6\% & 38.7\% & 46.1\% & 3.28 & 2\\ \midrule
    
    % 3.00%	14.10%	33.20%	49.70%	199.0046.1	34	165	3.296	2
    \multirow{2}{2em}{HPC0-D1} & F & 3.0\% & 14.1\% & 33.2\% & 49.7\% & 3.30 & 2\\ 
    % 3.10%	14.10%	38.20%	44.50%	192.00	33	159	3.239	2
    & M & 3.1\%  & 14.1\% & 38.2\% & 44.5\% & 3.24 & 2\\ \midrule
    
    % 9.60%	13.60%	29.80%	47.00%	199.00	46	153	3.142	2v
    \multirow{2}{2em}{HPC1-D0} & F & 9.6\% & 13.6\% & 29.8\% & 47.0\% & 3.14 & 2\\
    % 4.70%	18.80%	33.90%	42.70%	192.00	45	147	3.148	2
    & M & 4.7\%  & 18.8\% & 33.9\% & 42.7\% & 3.15 & 2\\ \midrule
    
    % 6.10%	10.10%	39.90%	43.90%	199.00	32	167	3.216	2
    \multirow{2}{2em}{HPC1-D1} & F & 6.1\% & 10.1\% & 39.9\% & 43.9\% & 3.22 & 2\\
    % 6.20%	17.20%	40.10%	36.50%	192.00	45	147	3.069	2
    & M & 6.2\%  & 17.2\% & 40.1\% & 36.5\% & 3.07 & 2\\ \midrule
    
    % 4.00%	16.10%	37.70%	42.20%	199.00	40	159	3.181	2
    \multirow{2}{2em}{HPC2-D0} & F & 4.0\% & 16.1\% & 37.7\% & 42.2\% & 3.18 & 2\\
    % 4.20%	16.70%	38.00%	41.10%	192.00	40	152	3.16	2
    & M & 4.2\%  & 16.7\% & 38.0\% & 41.1\% & 3.16 & 2\\ \midrule
    
    % 7.50%	14.60%	36.70%	41.20%	199.00	44	155	3.116	2
    \multirow{2}{2em}{HPC2-D1} & F & 7.5\% & 14.6\% & 36.7\% & 41.2\% & 3.12 & 2\\
    % 3.70%	14.70%	38.70%	42.90%	192.00	35	157	3.208	2
    & M & 3.7\%  & 14.7\% & 38.7\% & 42.9\% & 3.21 & 2\\ \midrule
    
    % 2.50%	9.00%	21.60%	66.80%	199.00	23	176	3.525	1
    \multirow{2}{2em}{HPC0-TTS} & F & 2.5\% & 9.0\% & 21.6\% & 66.8\% & 3.53 & 1\\
    % 1.80%	4.70%	37.10%	56.50%	192.00	12	180	3.485	1
    & M & 1.8\%  & 4.7\% & 37.1\% & 56.5\% & 3.49 & 1\\ 

    \bottomrule
  \end{tabular}
\vspace{-8mm}
\end{table}

{\em (ii) Quality \& naturalness} assessment is presented in Table~\ref{tab:mos} (with distinct tests for male \emph{M} and female \emph{F} target speakers). 
In addition to MOS scores (with \emph{PCM} recordings anchor) we provide 
the stimulus count (out of 40) with audible local problems that presumably have appeared as a result of forced-alignment errors (as tagged by a speech expert listening). We found the majority of such errors were too subtle to significantly reduce MOS, but may be indicative of potential sensitivity of a system to forced alignment during UPPT inference. The MOS results reveal that the quality of the proposed HPC-based prosody transfer systems is preserved as compared to the reference TTS operation (\emph{HPC0-TTS}) and outperforms the \emph{Ref} system. The MOS score differences between various HPC-based system configurations is found to be not statistically significant ($p=0.05$). However, based on the alignment-error counts, the models based on the phone-level HPCs (\emph{HPC2-D0},\emph{HPC2-D1}) are not recommended, although they result in the highest prosody similarity scores (Tables \ref{tab:male_prossim}, \ref{tab:female_prossim}).  

{\em (iii) Speaker similarity} evaluation results are finally presented in Table~\ref{tab:spksim} (with distinct tests for male \emph{M} and female \emph{F} target speakers, and all the systems compared to the regular TTS system \emph{HPC0-TTS}).  The rightmost column identifies
systems that do not differ significantly (in terms of Barnard's exact test) with the same rank, demonstrating for all the HPC-based systems non-significantly different speaker similarity scores
\footnote{Audio samples are available at \href{https://ibm.biz/IS23-TBE}{https://ibm.biz/IS23-TBE}}.

\section{Summary}
We presented a novel HPC-based neural TTS system with UPPT functionality,
and demonstrated through extensive perceptual evaluations that the 
systems can transfer prosody from input exemplars uttered by novel speakers to various trained TTS voices with high precision while incurring no quality 
degradation and preserving the target speaker similarity. 
Extensions of this work will dive deeper into the robustness of the
techniques under more extreme transfer conditions (e.g., unusually elongated
sounds), and look into going beyond pitch and duration to 
transfer various timbral effects observed in an input recording, as might
be the case with highly emotive speech. 
% Note: Do not speculate HOW we might solve the problem. We don't want to
% reveal anything :) -- RF

\bibliographystyle{IEEEtran}
\bibliography{IS23_TuningbyEx}
%\bibliography{C://Papers//2023//IS2023_TbyEx.arXiv//learning,C://Papers//2023//IS2023_TbyEx.arXiv//nn,C://Papers//2023//IS202_TbyEx.arXiv//nlp,C://Papers//2023//IS2023_TbyEx.arXiv//optimization,C://Papers//2023//IS2023_TbyEx.arXiv//prosody,C://Papers//2023//IS2023_TbyEx.arXiv//speech,C://Papers//2023//IS2023_TbyEx.arXiv//tts,C://Papers//2023//IS2023_TbyEx.arXiv//ttse2e}

\end{document}